\documentclass[aps,prl,twocolumn,superscriptaddress,floatfix]{revtex4-1}
\usepackage{graphicx,amsmath,amssymb,subfigure,epsfig,psfrag,verbatim,multirow}
\usepackage[normalem]{ulem}  
\usepackage[usenames,dvipsnames,sgvnames,table]{xcolor}  

\begin{document}
\title{Classifying surface probe images in strongly correlated electronic systems via machine learning}

\author{L. Burzawa}
\affiliation{Department of Physics and Astronomy, Purdue University, West Lafayette, IN 47907, USA}
\author{S. Liu}
\affiliation{Department of Physics and Astronomy, Purdue University, West Lafayette, IN 47907, USA}
\author{E. W. Carlson}
\email{ewcarlson@purdue.edu}
\affiliation{Department of Physics and Astronomy, Purdue University, West Lafayette, IN 47907, USA}

\date{\today}

\begin{abstract}
Scanning probe experiments such as scanning tunneling microscopy (STM) and atomic force microscopy (AFM) on strongly correlated electronic systems often reveal complex pattern formation on multiple length scales.\cite{basov-science,kohsaka-science-2007} By studying the universal scaling in these images, we have shown in several distinct correlated electronic systems that the pattern formation is driven by proximity to a disorder-driven critical point,\cite{phillabaum-2012,shuo-vo2} 
revealing a unification of the pattern formation in these materials.  
As an alternative approach to this image classification problem of novel materials, here we report the first investigation of the machine learning method to determine which underlying physical model is driving pattern formation in a system. Using a neural network architecture, we are able to achieve 97$\%$ accuracy on classifying configuration images from three models with Ising symmetry. 
This investigation also demonstrates that machine learning can capture the implicit universal behavior of a physical system. This broadens our understanding of what machine learning can do, and we expect more synergy between machine learning and condensed matter physics in the future.

\end{abstract}
\maketitle

\section{Introduction}

Scanning probe experiments often reveal complex pattern formation at
the surface of strongly correlated electronic systems\cite{basov-science,kohsaka-science-2007,phillabaum-2012,shuo-vo2,Dagotto,
dagotto-moreo-manganite}. Since their invention in 1982, scanning
probes have revolutionized our understanding of materials, 
yielding an ever increasing wealth of data on a wide variety of 
materials\cite{stm-rmp-nobel}. To date, the majority of theoretical
treatments have focused on microscopic physics \cite{basov-rmp},
with few theoretical treatments offering guidance for how to
interpret the detailed spatial information available in the emergent
multiscale pattern formation often observed on surfaces. 
For systems 
near criticality, the spatial configurations of geometric
clusters become scale-free, displaying spatial complexity on multiple length scales
in a way that is controlled by the critical fixed point. 
Therefore,
the geometric properties encode critical exponents,
as we have shown elsewhere.\cite{phillabaum-2012,superstripes-erice-2014,shuo-vo2} 
Such scale-free complexity can arise from interactions deep
within a material, or from surface-only physics, or even
from a non-interacting model at the right concentration.  
We show here that artificial neural networks 
can be trained to identify which physics is responsible 
for the complex pattern formation, identifying whether
interactions are present, and if so, whether the interactions
arise from deep inside the material, or whether they
arise from surface physics.

Machine learning (ML) 
is a burgeoning field in computer science and data
science, with broad applications across disciplines such as bioinformatics, computer vision,
marketing, economics, and medical diagnosis. 
A computer program is said to exhibit ML if its ability to perform a given task
increases with experience, as determined by some performance metric. 
That is, the program accumulates iterative modifications when presented with
certain input (``experience''), in such a way as to improve its performance
at the task, without those modifications being explicitly programmed.\cite{ML-def}

In physics, ML has been applied to to physics at a range of scales,
from  galaxy clusters\cite{astro-ml} to elementary particles\cite{particle_phys-ml}.
ML is beginning to be applied in condensed matter physics,
for example in quantum many body problems\cite{troyer-ml},  electronic quantum transport\cite{lopez-ml}, glassy dynamics\cite{schoenholz-ml}, phase transitions\cite{wang-ml,melko-ml}, renormalization group\cite{ML-prec4,ML-prec5}, and big data issues of  materials science\cite{kusne-ml,ghiringhelli-ml,kalinin-ml}.
Arsenault {\em et al.} have used ML to address problems in many-body physics such as the Anderson impurity model\cite{millis-prb} 
and dynamical mean-field theory.\cite{millis-arxiv}.
Carleo and Troyer used ML to study the wavefunction of quantum many-body 
interacting spin systems.\cite{troyer-ml} 
Wang showed that an unsupervised learning algorithm could ``discover''
that the order parameter and structure factor change significantly as the temperature
is varied through the phase transition.\cite{wang-ml} 
Carrasquilla and Melko have shown that for a few different models, supervised learning
can distinguish the ordered from the disordered phase.\cite{melko-ml}

However, no one has yet tasked ML with identifying
which underlying Hamiltonian is actually responsible for the phase transition.  
We show in this paper that ML can  be used to identify which model
was used to generate a particular spin configuration, focusing on configurations
that are close to criticality.   We focus on near-critical configurations, since
these are most relevant to interpreting the multiscale pattern formation observed
in many spatially resolved experiments.\cite{basov-science,kohsaka-science-2007,phillabaum-2012,shuo-vo2,Dagotto,
dagotto-moreo-manganite}
This type of identification can reveal information 
such as which interactions are important, 
whether quenched  disorder is a relevant term in the Hamiltonian, 
and how many dimensions are involved in the phenomenon
(to determine, for example, whether observed patterns in data
are driven by surface physics, or from the bulk of the material).

\begin{figure}
\centerline{\includegraphics[width=.95\columnwidth]{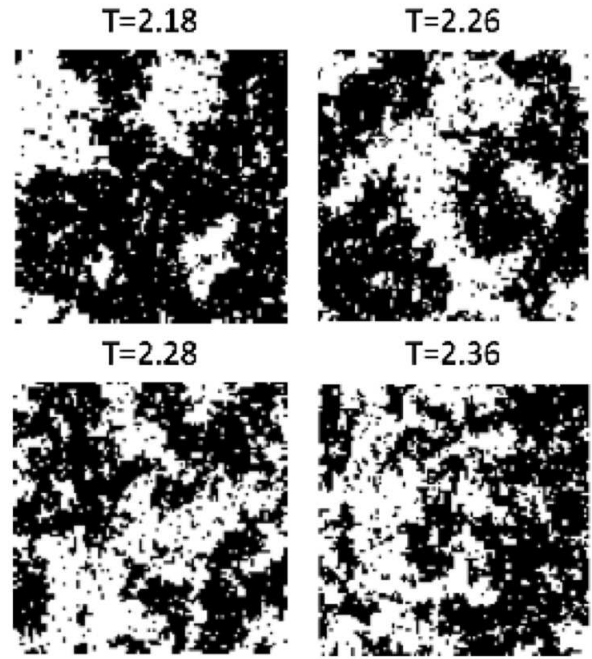}}
\caption{Ising 2D images. Row 1: $T\textless T_c$,  row 2: $T\approx T_c=2.269$, row 3: $T\textgreater T_c$.  Temperatures are in units of J, which is the coupling strength between Ising variables.
Black and white pixels represent Ising variables $\sigma = +1$ and $-1$ respectively.
}
\label{ising2D}
\end{figure}
\begin{figure}
\centerline{\includegraphics[width=.95\columnwidth]{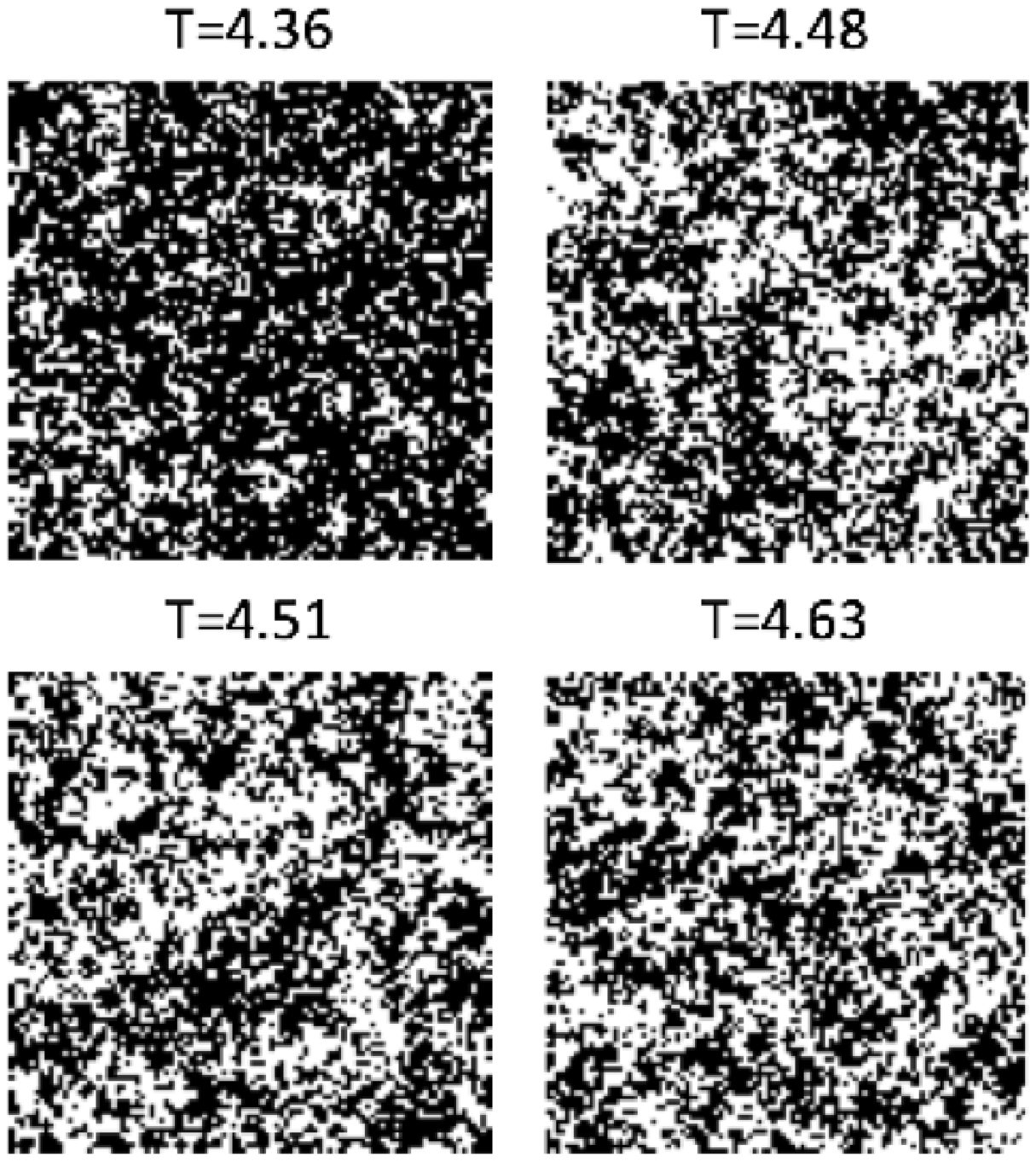}}
\caption{Ising 3D images. Row 1: $T\textless T_c$,  row 2: $T\approx T_c=4.512$, row 3: $T\textgreater T_c$.  Temperatures are in units of J, which is the coupling strength between Ising variables.
Black and white pixels represent Ising variables $\sigma = +1$ and $-1$ respectively.
}
\label{ising3D}
\end{figure}
\begin{figure}
\centerline{\includegraphics[width=.95\columnwidth]{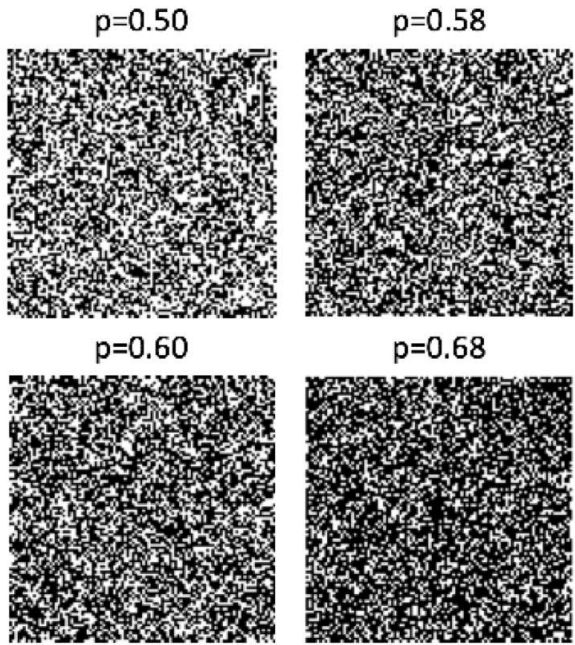}}
\caption{Percolation images. Row 1: $p\textless p_c$, row 2: $p\approx p_c$, row 3: $p\textgreater p_c$.  On a square lattice, the site percolation threshold is $p_c = 0.59$.
Black and white pixels represent variables $\sigma = +1$ and $-1$ respectively.
}
\label{percolation}
\end{figure}

In this paper, we are interested in whether ML can
classify microscopy images according to the underlying physics
driving the complex pattern formation. To be more specific, we want
to know whether ML can capture the universal
properties and critical behavior implied by the image patterns, and
then identify which model generated the image.  
In the present work we
limit ourselves to three theoretical models: the two-dimensional (2D) clean Ising model on a square lattice (Fig.~\ref{ising2D});  
the three-dimensional (3D) clean Ising model on a cubic lattice (Fig.~\ref{ising3D}); 
and the square lattice site percolation model  (Fig.~\ref{percolation}). 
The Ising models may be written as: 
\begin{equation}
H=- J \sum_{<i j>} \sigma_i \sigma_j 
\end{equation}
where $\sigma_i = \pm 1$ and $J$ is the coupling strength between nearest neighbor sites,
and the summation runs over the sites of either a two-dimensional square lattice 
or a three-dimensional cubic lattice.  

Ising models were first used to describe magnetic transitions, 
from a ferromagnetically ordered phase at low temperature $T < T_c$,
to a paramagnetic, disordered phase 
above the magnetic ordering temperature $T_c$,\cite{fisher-rmp,stanley-book}  
in systems where magnetic moments are constrained to point 
either ``up'' or ``down.''
However, the model can be applied to a wide variety of physical systems.
For example, the behavior of the critical endpoint of the
liquid-gas phase transition is well described by the criticality
of the three-dimensional Ising model.  
Electrons inside of solids have their own phase transitions, and 
when comparing to experiments on these systems,
$\sigma$ may be viewed as a generalized pseudospin in the 
context of two-component electronic behavior.  
This may be mapped, for example, to  the two perpendicular orientations of nematic stripes in cuprate superconductors,\cite{phillabaum-2012}
or to the metal and insulator islands in VO$_2$.\cite{shuo-vo2}

In the absence of interactions, pseudospins may be said to follow a percolation model,
which is a non-interacting model.
Therefore we also consider site percolation on a square lattice, assigning a probability $p$ 
to having pseudospin $\sigma =1$ on any given site, otherwise
the pseudospin is assigned as $\sigma = -1$.  
In two dimensions, this model has a continuous phase transition (and therefore
exhibits criticality) at $p_c = 0.59$.  
We aim to explore 
the efficiency of ML for capturing features associated with
the corresponding universality class, including interactions, disorder, and dimension.
Other universal features, such as the type of random disorder in the system, and the symmetry
of the order parameter, are left for future work.  

\section{Methods}
With ML, a software program undergoes significant changes based on new input,
without those changes being explicitly hard-coded by the programmer. 
Rather, training algorithms contained within the program train
the neural network as new input is received.\cite{ML-def}
Whereas human visual pattern classification and recognition can be viewed as a qualitative process, ML  turns this into an explicitly quantitative process, albeit within some margin of error. 
In our case, we have used a scaled conjugate gradient (SCG) algorithm\cite{moller-scg}
under supervised learning conditions 
 to train a neural network algorithm
using the MATLAB Neural Network Toolbox through XSEDE\cite{XSEDE}. 
With supervised learning, the program is presented with a training set of data for which
the ``right answer'' is also supplied to the training algorithm.  
The point is to develop an algorithm that gives the correct output for a given input.
In our case, the goal will be to identify the underlying interacting physics model (the output) from which 
a particular Ising spin configuration was generated (the input).  

\begin{figure}
\centerline{\includegraphics[width=.8\columnwidth]{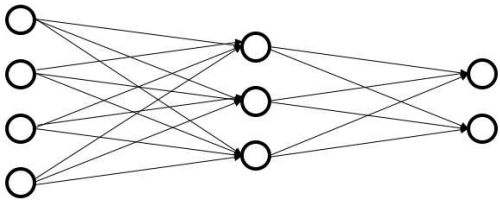}}
\caption{
Schematic of a small artificial neural network. In this case there are 4 features in the input data (for example pixels in a 2x2 image), 3 hidden neurons, and 2 outputs (for example cat/dog, or percolation/Ising).
}
\label{fig-dac}
\end{figure}

In order to accomplish this, we have used an artificial neural network with a single hidden layer containing 200 ``neurons''. 
Each neuron consists of multiple inputs, a nonlinear activation function, and a single output.
We use a hyperbolic tangent activation function, which results in
outputs between -1 and 1. 
The network is defined by its topology (how the neurons are interconnected) and the weights associated with each connection.  
We use a feedforward neural network, which 
can be represented by a directed acyclic graph where each edge has a weight and each node has a bias. 
A very small example of a neural network topology is illustrated in Fig.~\ref{fig-dac}.  The input layer consists 
of the black and white image itself, where each circle represents one pixel. 
In the hidden layer, each circle represents a single neuron, with multiple inputs,
which turns on according to some nonlinear function of its input weights.  
The output layer in our case has three nodes, one for
each of the three models which may generate a spatially complex image
near criticality.  

The goal of neural network training is to optimize the weights and biases of the network,
by minimizing a loss function.  
In the present case, we use cross entropy as the loss function. 
The training proceeds towards a point where the loss function is at a minimum.
We have used scaled conjugate gradient descent method\cite{moller-scg}  to
minimize the loss function.  
We furthermore use backpropagation of errors, a type of automatic differentiation in which
the derivatives associated with the chain rule are computed in reverse order ({\em i.e.} working from
the topmost dependent variable down through to the independent variables).

\begin{table}[htb] 
\centering
\begin{tabular}{|c|c|c|c|c|c|c|c|}
\hline
Neurons & 50 & 75 & 100 & 125 & 150 & 200 & 250 \\ 
\hline
\% Accuracy & 94.93 & 95.53 & 95.95 & 96.48 & 96.68 & 97.00 & 96.72 \\
\hline
\end{tabular}
\caption{Classification accuracy for a given number of hidden neurons. 
Accuracy initially increases with increasing number of hidden neurons.}
\label{tab:neurons}
\end{table}

We use the 2D percolation model and Monte Carlo simulations of the 2D clean Ising model and 3D clean Ising model  to generate images to train the neural network.  For the 2D models, the neural network is directly fed spin configurations near criticality for training purposes.  For the 3D model, the neural network is fed spin configurations from a 2D {\em slice} of the 3D lattice.  This is to mimic surface probe experiments, which have access to only 2D information.
 The goal is for the neural network to learn to identify which model generated which images.  

The ``interesting'' cases occur near criticality, where geometric clusters (as defined by connected sets of like-spin nearest neighbors) 
grow in such a way as to have structure over multiple length scales.
Near criticality, systems experience fluctuations over all length scales from 
atomic to the size of the system.    Because there is no characteristic length scale in between,
critical systems are described by power law behavior,
where the exponent of each power law is called a ``critical exponent.''  
At any second order phase transition, the unique set of critical exponents
acts like a fingerprint to identify the universality class of the phase transition,
which is set by universal features such as the dimension of the lattice and the symmetry of the order parameter
($Z_2$ in the case of Ising variables $\sigma = \pm 1$), and independent of
``non-universal'' short-distance physics like the inclusion of 2nd-nearest-neighbor interactions.\cite{fisher-rmp,stanley-book}
Ultimately, this behavior arises because the geometric clusters have a fractal character near criticality,
whereby each cluster is scale-free.
In these cases, we have shown\cite{phillabaum-2012,superstripes-erice-2014,shuo-vo2}  that the statistics of the shapes of the clusters in the images encodes the universality class of the underlying model.
The hardest cases to judge by eye are those with roughly equal numbers of up and down spins.  
For this reason, we focus attention within 4\% of the critical region of the Ising models
(since magnetization rapidly develops inside the ordered region), and within 15\% of the percolation model
(because equal numbers of up and down spins in that case is p=0.5, which is close but not at the critical
value of $p_c = 0.59$ for site percolation on a square lattice).  
Note that net magnetization is not a trivial discriminator of these models near criticality.
Within these parameter ranges, each model covers states with no net magnetization
as well as those with net magnetization in the thermodynamic (infinite size) limit.
Even more importantly, for the same nominal set of model parameters,
thermal fluctuations can cause a net magnetization to appear in a finite size system 
for $T > T_c$. 
It is especially important  to distinguish whether interactions are present, 
in order to determine in any given surface probe experiment whether interactions 
are responsible for the observed pattern formation, in order to distinguish it from
putative ``dirt effects'' which may arise at the surface of a material.  
Because the Ising models are near criticality, we use the Wolff algorithm, and sample spin configurations every 100 Monte Carlo steps to ensure independent sampling.  
We simulate system sizes of $L^2 = 100^2$ for the 2D case, and $L^3 = 100^3$ for the 3D  case.
(For comparison, experimental images from surface probes in condensed matter physics today are typically about 
$200 \times 200$  pixels.\cite{basov-rmp})
The images passed to the neural network consist of $100 \times 100 = 10,000$ pixels, called ``features'' in the context of ML;
the number of training examples given to the neural network must be significantly greater than 
We use 400,000 configurations from each model, for a total of 1,200,000 configurations.
These configurations are then divided randomly, using 70\% for training, 
15\% for validation, and 15\% for testing.

\section{Results and Discussion}
We find that the neural network, once trained to identify which model produced which spin configurations,
achieves an average classification accuracy 
of about  97\%, once the number of hidden neurons is at least 150. 
The dependence of the classification accuracyon the number of hidden neurons is shown in  Table~\ref{tab:neurons}.
The error bars are calculated from the case with 200 neurons where multiple training runs where conducted. The classification error drops from around 5$\%$ at 50 neurons to around 3$\%$ at 150 neurons. Results between 150 and 250 neurons are very similar and adding more neurons than 150 does not help much, although there does appear to be a minimum in the error for 200 neurons.

The case study presented in Table~\ref{tab:error-matrix} shows how errors were distributed among the models. The rows represent the models that images belong to and columns represent the models that images were classified as. The most misclassified model is the 3D Ising model. 
It was particularly hard for the neural network to distinguish between the 2D and 3D Ising models.  
Ising 2D and percolation are the most easily distinguished with almost negligible errors.

We have shown that ML can be used to classify configuration images into their associated universality classes, which means that it identifies implicit universal properties  underlying the complex pattern formation of image configurations near a critical point.  The conventional approach to identifying a universality class is to first explicitly fit the experimental data in order to find the critical exponents, then match that set of critical exponents to a known universality class. 
However, the approach developed here does not depend upon critical exponents. Instead, we feed image data and universality class labels to the machine learning model and let the ML algorithm develop its own internal parameters in the neural network. We haven't programmed the model to include any theories from physics, but the model learns directly from the data, and then captures the universal features in the configuration images to identify the 
universality class. 
Thus, we have shown that it is possible to go directly from data to an identification of the universality class of that data, without explicitly considering any critical exponents or  scaling theory. 

This finding broadens our understanding of the extent to which ML can be applied, 
and thus has important implications in the field of ML and computer vision where 
the image classification problem is mainly focused on explicit object recognition, 
rather than discovering underlying physics.  We also expect 
that machine learning can be used to study and even learn more complex physics given large 
enough datasets. This synergy between machine learning and physics opens a new perspective 
for future research in condensed matter physics, or even in all fields of physics, that we can 
draw important conclusions via ML,  {\em without the need to interpret the intermediate steps} of the ML algorithm.

\begin{table}
\begin{center}
\begin{tabular}{cc|c|c|c|}
\cline{3-5}
& & \multicolumn{3}{ c| }{Neural network output} \\
\cline{3-5}
& &  perc & I-2D & I-3D \\ 
\hline
\multicolumn{1}{ |c  }{\multirow{3}{*}{Input model}} & \multicolumn{1}{ |c|  }{perc} & 98.05\% & 0.01\% & 1.94\% \\
\cline{2-5} 
\multicolumn{1}{ |c  }{} &\multicolumn{1}{ |c|  }{ I-2D} & 0.03\% & 98.30\% & 1.67\% \\
\cline{2-5}
\multicolumn{1}{ |c  }{} & \multicolumn{1}{ |c|  }{I-3D} & 1.82\% & 2.96\% & 95.22\% \\
\hline
\end{tabular}
\caption{Classfication results for testing data 
for the case of 200 neurons in the hidden layer.  
The model listed on each row denotes the Hamiltonian used to generate spin configurations to
present to the neural network.  In each column, the output of the neural network is listed.  
The goal is to train the neural network to identify which model produced which spin configurations.
}
\end{center}
\label{tab:error-matrix}
\end{table}

\section {Conclusion}
In conclusion, we have shown that machine learning  can successfully determine 
which spin configurations were generated from which  theoretical model when the images are near criticality.   As scanning probe experimental capabilities grow, they are producing a growing wealth of data, including more examples of systems in which the electronic textures display multiscale pattern formation at the surface of the material.\cite{basov-science,kohsaka-science-2007,V2O3,milan-iridates}   
In cases where the pattern formation is driven by proximity to a critical point,\cite{phillabaum-2012,superstripes-erice-2014,shuo-vo2}  the techniques employed here can be used to identify the underlying 
physics 
driving the pattern formation, without
the need to explicitly determine critical exponents or scaling forms.


\section{Acknowledgements}
LB acknowledges support from the SURF program through the College of Engineering at Purdue University. 
SL acknowledges support from the Bilsland Dissertation Fellowship at Purdue University. 
EWC and SL acknowledge support from NSF DMR-1508236 and 
Dept. of Education Grant No. P116F140459.   
This work used the Extreme Science and Engineering Discovery Environment (XSEDE), which is supported by National Science Foundation grant number ACI-1053575.


\begin{thebibliography}{30}%
\makeatletter
\providecommand \@ifxundefined [1]{%
 \@ifx{#1\undefined}
}%
\providecommand \@ifnum [1]{%
 \ifnum #1\expandafter \@firstoftwo
 \else \expandafter \@secondoftwo
 \fi
}%
\providecommand \@ifx [1]{%
 \ifx #1\expandafter \@firstoftwo
 \else \expandafter \@secondoftwo
 \fi
}%
\providecommand \natexlab [1]{#1}%
\providecommand \enquote  [1]{``#1''}%
\providecommand \bibnamefont  [1]{#1}%
\providecommand \bibfnamefont [1]{#1}%
\providecommand \citenamefont [1]{#1}%
\providecommand \href@noop [0]{\@secondoftwo}%
\providecommand \href [0]{\begingroup \@sanitize@url \@href}%
\providecommand \@href[1]{\@@startlink{#1}\@@href}%
\providecommand \@@href[1]{\endgroup#1\@@endlink}%
\providecommand \@sanitize@url [0]{\catcode `\\12\catcode `\$12\catcode
  `\&12\catcode `\#12\catcode `\^12\catcode `\_12\catcode `\%12\relax}%
\providecommand \@@startlink[1]{}%
\providecommand \@@endlink[0]{}%
\providecommand \url  [0]{\begingroup\@sanitize@url \@url }%
\providecommand \@url [1]{\endgroup\@href {#1}{\urlprefix }}%
\providecommand \urlprefix  [0]{URL }%
\providecommand \Eprint [0]{\href }%
\providecommand \doibase [0]{http://dx.doi.org/}%
\providecommand \selectlanguage [0]{\@gobble}%
\providecommand \bibinfo  [0]{\@secondoftwo}%
\providecommand \bibfield  [0]{\@secondoftwo}%
\providecommand \translation [1]{[#1]}%
\providecommand \BibitemOpen [0]{}%
\providecommand \bibitemStop [0]{}%
\providecommand \bibitemNoStop [0]{.\EOS\space}%
\providecommand \EOS [0]{\spacefactor3000\relax}%
\providecommand \BibitemShut  [1]{\csname bibitem#1\endcsname}%
\let\auto@bib@innerbib\@empty
\bibitem [{\citenamefont {Qazilbash}\ \emph {et~al.}(2007)\citenamefont
  {Qazilbash}, \citenamefont {Brehm}, \citenamefont {Chae}, \citenamefont {Ho},
  \citenamefont {Andreev}, \citenamefont {Kim}, \citenamefont {Yun},
  \citenamefont {Balatsky}, \citenamefont {Maple}, \citenamefont {Keilmann},
  \citenamefont {Kim},\ and\ \citenamefont {Basov}}]{basov-science}%
  \BibitemOpen
  \bibfield  {author} {\bibinfo {author} {\bibfnamefont {M.~M.}\ \bibnamefont
  {Qazilbash}}, \bibinfo {author} {\bibfnamefont {M.}~\bibnamefont {Brehm}},
  \bibinfo {author} {\bibfnamefont {B.-G.}\ \bibnamefont {Chae}}, \bibinfo
  {author} {\bibfnamefont {P.-C.}\ \bibnamefont {Ho}}, \bibinfo {author}
  {\bibfnamefont {G.~O.}\ \bibnamefont {Andreev}}, \bibinfo {author}
  {\bibfnamefont {B.-J.}\ \bibnamefont {Kim}}, \bibinfo {author} {\bibfnamefont
  {S.~J.}\ \bibnamefont {Yun}}, \bibinfo {author} {\bibfnamefont {A.~V.}\
  \bibnamefont {Balatsky}}, \bibinfo {author} {\bibfnamefont {M.~B.}\
  \bibnamefont {Maple}}, \bibinfo {author} {\bibfnamefont {F.}~\bibnamefont
  {Keilmann}}, \bibinfo {author} {\bibfnamefont {H.-T.}\ \bibnamefont {Kim}}, \
  and\ \bibinfo {author} {\bibfnamefont {D.~N.}\ \bibnamefont {Basov}},\ }\href
  {\doibase 10.1126/science.1150124} {\bibfield  {journal} {\bibinfo  {journal}
  {Science}\ }\textbf {\bibinfo {volume} {318}},\ \bibinfo {pages} {1750}
  (\bibinfo {year} {2007})}\BibitemShut {NoStop}%
\bibitem [{\citenamefont {Kohsaka}\ \emph {et~al.}(2007)\citenamefont
  {Kohsaka}, \citenamefont {Taylor}, \citenamefont {Fujita}, \citenamefont
  {Schmidt}, \citenamefont {Lupien}, \citenamefont {Hanaguri}, \citenamefont
  {Azuma}, \citenamefont {Takano}, \citenamefont {Eisaki}, \citenamefont
  {Takagi}, \citenamefont {Uchida},\ and\ \citenamefont
  {Davis}}]{kohsaka-science-2007}%
  \BibitemOpen
  \bibfield  {author} {\bibinfo {author} {\bibfnamefont {Y.}~\bibnamefont
  {Kohsaka}}, \bibinfo {author} {\bibfnamefont {C.}~\bibnamefont {Taylor}},
  \bibinfo {author} {\bibfnamefont {K.}~\bibnamefont {Fujita}}, \bibinfo
  {author} {\bibfnamefont {A.}~\bibnamefont {Schmidt}}, \bibinfo {author}
  {\bibfnamefont {C.}~\bibnamefont {Lupien}}, \bibinfo {author} {\bibfnamefont
  {T.}~\bibnamefont {Hanaguri}}, \bibinfo {author} {\bibfnamefont
  {M.}~\bibnamefont {Azuma}}, \bibinfo {author} {\bibfnamefont
  {M.}~\bibnamefont {Takano}}, \bibinfo {author} {\bibfnamefont
  {H.}~\bibnamefont {Eisaki}}, \bibinfo {author} {\bibfnamefont
  {H.}~\bibnamefont {Takagi}}, \bibinfo {author} {\bibfnamefont
  {S.}~\bibnamefont {Uchida}}, \ and\ \bibinfo {author} {\bibfnamefont {J.~C.}\
  \bibnamefont {Davis}},\ }\href@noop {} {\bibfield  {journal} {\bibinfo
  {journal} {Science}\ }\textbf {\bibinfo {volume} {315}},\ \bibinfo {pages}
  {1380} (\bibinfo {year} {2007})}\BibitemShut {NoStop}%
\bibitem [{\citenamefont {Phillabaum}\ \emph {et~al.}(2012)\citenamefont
  {Phillabaum}, \citenamefont {Carlson},\ and\ \citenamefont
  {Dahmen}}]{phillabaum-2012}%
  \BibitemOpen
  \bibfield  {author} {\bibinfo {author} {\bibfnamefont {B.}~\bibnamefont
  {Phillabaum}}, \bibinfo {author} {\bibfnamefont {E.~W.}\ \bibnamefont
  {Carlson}}, \ and\ \bibinfo {author} {\bibfnamefont {K.~A.}\ \bibnamefont
  {Dahmen}},\ }\href@noop {} {\bibfield  {journal} {\bibinfo  {journal} {Nat.
  Commun.}\ }\textbf {\bibinfo {volume} {3}},\ \bibinfo {pages} {915} (\bibinfo
  {year} {2012})}\BibitemShut {NoStop}%
\bibitem [{\citenamefont {Liu}\ \emph {et~al.}(2016)\citenamefont {Liu},
  \citenamefont {Phillabaum}, \citenamefont {Carlson}, \citenamefont {Dahmen},
  \citenamefont {Vidhyadhiraja}, \citenamefont {Qazilbash},\ and\ \citenamefont
  {Basov}}]{shuo-vo2}%
  \BibitemOpen
  \bibfield  {author} {\bibinfo {author} {\bibfnamefont {S.}~\bibnamefont
  {Liu}}, \bibinfo {author} {\bibfnamefont {B.}~\bibnamefont {Phillabaum}},
  \bibinfo {author} {\bibfnamefont {E.~W.}\ \bibnamefont {Carlson}}, \bibinfo
  {author} {\bibfnamefont {K.~A.}\ \bibnamefont {Dahmen}}, \bibinfo {author}
  {\bibfnamefont {N.~S.}\ \bibnamefont {Vidhyadhiraja}}, \bibinfo {author}
  {\bibfnamefont {M.~M.}\ \bibnamefont {Qazilbash}}, \ and\ \bibinfo {author}
  {\bibfnamefont {D.~N.}\ \bibnamefont {Basov}},\ }\href@noop {} {\bibfield
  {journal} {\bibinfo  {journal} {Phys. Rev. Lett.}\ }\textbf {\bibinfo
  {volume} {116}},\ \bibinfo {pages} {036401} (\bibinfo {year}
  {2016})}\BibitemShut {NoStop}%
\bibitem [{\citenamefont {Dagotto}(2005)}]{Dagotto}%
  \BibitemOpen
  \bibfield  {author} {\bibinfo {author} {\bibfnamefont {E.}~\bibnamefont
  {Dagotto}},\ }\href {\doibase 10.1126/science.1107559} {\bibfield  {journal}
  {\bibinfo  {journal} {Science}\ }\textbf {\bibinfo {volume} {309}},\ \bibinfo
  {pages} {257} (\bibinfo {year} {2005})}\BibitemShut {NoStop}%
\bibitem [{\citenamefont {Moreo}\ \emph {et~al.}(2000)\citenamefont {Moreo},
  \citenamefont {Mayr}, \citenamefont {Feiguin}, \citenamefont {Yunoki},\ and\
  \citenamefont {Dagotto}}]{dagotto-moreo-manganite}%
  \BibitemOpen
  \bibfield  {author} {\bibinfo {author} {\bibfnamefont {A.}~\bibnamefont
  {Moreo}}, \bibinfo {author} {\bibfnamefont {M.}~\bibnamefont {Mayr}},
  \bibinfo {author} {\bibfnamefont {A.}~\bibnamefont {Feiguin}}, \bibinfo
  {author} {\bibfnamefont {S.}~\bibnamefont {Yunoki}}, \ and\ \bibinfo {author}
  {\bibfnamefont {E.}~\bibnamefont {Dagotto}},\ }\href {\doibase
  10.1103/PhysRevLett.84.5568} {\bibfield  {journal} {\bibinfo  {journal}
  {Phys. Rev. Lett.}\ }\textbf {\bibinfo {volume} {84}},\ \bibinfo {pages}
  {5568} (\bibinfo {year} {2000})}\BibitemShut {NoStop}%
\bibitem [{\citenamefont {Binnig}\ and\ \citenamefont
  {Rohrer}(2013)}]{stm-rmp-nobel}%
  \BibitemOpen
  \bibfield  {author} {\bibinfo {author} {\bibfnamefont {G.}~\bibnamefont
  {Binnig}}\ and\ \bibinfo {author} {\bibfnamefont {H.}~\bibnamefont
  {Rohrer}},\ }\href@noop {} {\bibfield  {journal} {\bibinfo  {journal}
  {Reviews of Modern Physics}\ }\textbf {\bibinfo {volume} {71}},\ \bibinfo
  {pages} {324} (\bibinfo {year} {2013})}\BibitemShut {NoStop}%
\bibitem [{\citenamefont {Bonnell}\ \emph {et~al.}(2012)\citenamefont
  {Bonnell}, \citenamefont {Basov}, \citenamefont {Bode}, \citenamefont
  {Diebold}, \citenamefont {Kalinin}, \citenamefont {Madhavan}, \citenamefont
  {Novotny}, \citenamefont {Salmeron}, \citenamefont {Schwarz},\ and\
  \citenamefont {Weiss}}]{basov-rmp}%
  \BibitemOpen
  \bibfield  {author} {\bibinfo {author} {\bibfnamefont {D.~A.}\ \bibnamefont
  {Bonnell}}, \bibinfo {author} {\bibfnamefont {D.~N.}\ \bibnamefont {Basov}},
  \bibinfo {author} {\bibfnamefont {M.}~\bibnamefont {Bode}}, \bibinfo {author}
  {\bibfnamefont {U.}~\bibnamefont {Diebold}}, \bibinfo {author} {\bibfnamefont
  {S.~V.}\ \bibnamefont {Kalinin}}, \bibinfo {author} {\bibfnamefont
  {V.}~\bibnamefont {Madhavan}}, \bibinfo {author} {\bibfnamefont
  {L.}~\bibnamefont {Novotny}}, \bibinfo {author} {\bibfnamefont
  {M.}~\bibnamefont {Salmeron}}, \bibinfo {author} {\bibfnamefont {U.~D.}\
  \bibnamefont {Schwarz}}, \ and\ \bibinfo {author} {\bibfnamefont {P.~S.}\
  \bibnamefont {Weiss}},\ }\href@noop {} {\bibfield  {journal} {\bibinfo
  {journal} {Reviews of Modern Physics}\ }\textbf {\bibinfo {volume} {84}},\
  \bibinfo {pages} {1343} (\bibinfo {year} {2012})}\BibitemShut {NoStop}%
\bibitem [{\citenamefont {Carlson}\ \emph {et~al.}(2015)\citenamefont
  {Carlson}, \citenamefont {Liu}, \citenamefont {Phillabaum},\ and\
  \citenamefont {Dahmen}}]{superstripes-erice-2014}%
  \BibitemOpen
  \bibfield  {author} {\bibinfo {author} {\bibfnamefont {E.~W.}\ \bibnamefont
  {Carlson}}, \bibinfo {author} {\bibfnamefont {S.}~\bibnamefont {Liu}},
  \bibinfo {author} {\bibfnamefont {B.}~\bibnamefont {Phillabaum}}, \ and\
  \bibinfo {author} {\bibfnamefont {K.~A.}\ \bibnamefont {Dahmen}},\
  }\href@noop {} {\bibfield  {journal} {\bibinfo  {journal} {Journal of
  Superconductivity and Novel Magnetism}\ } (\bibinfo {year}
  {2015})}\BibitemShut {NoStop}%
\bibitem [{\citenamefont {Russell}\ and\ \citenamefont
  {Norvig}(2015)}]{ML-def}%
  \BibitemOpen
  \bibfield  {author} {\bibinfo {author} {\bibfnamefont {S.}~\bibnamefont
  {Russell}}\ and\ \bibinfo {author} {\bibfnamefont {P.}~\bibnamefont
  {Norvig}},\ }\enquote {\bibinfo {title} {Artificial intelligence: A modern
  approach},}\ \ (\bibinfo  {publisher} {Pearson},\ \bibinfo {year} {2015})\
  Chap.~\bibinfo {chapter} {V}\BibitemShut {NoStop}%
\bibitem [{\citenamefont {Ball}\ and\ \citenamefont
  {Brunner}(2010)}]{astro-ml}%
  \BibitemOpen
  \bibfield  {author} {\bibinfo {author} {\bibfnamefont {N.}~\bibnamefont
  {Ball}}\ and\ \bibinfo {author} {\bibfnamefont {R.}~\bibnamefont {Brunner}},\
  }\href@noop {} {\bibfield  {journal} {\bibinfo  {journal} {Int. Journal of
  Modern Physics D}\ }\textbf {\bibinfo {volume} {19}},\ \bibinfo {pages}
  {1049} (\bibinfo {year} {2010})}\BibitemShut {NoStop}%
\bibitem [{\citenamefont {Whiteson}\ and\ \citenamefont
  {Whiteson}(2009)}]{particle_phys-ml}%
  \BibitemOpen
  \bibfield  {author} {\bibinfo {author} {\bibfnamefont {S.}~\bibnamefont
  {Whiteson}}\ and\ \bibinfo {author} {\bibfnamefont {D.}~\bibnamefont
  {Whiteson}},\ }\href@noop {} {\bibfield  {journal} {\bibinfo  {journal}
  {Engineering Applications of AI}\ }\textbf {\bibinfo {volume} {22}},\
  \bibinfo {pages} {1203} (\bibinfo {year} {2009})}\BibitemShut {NoStop}%
\bibitem [{\citenamefont {Carleo}\ and\ \citenamefont
  {Troyer}(2017)}]{troyer-ml}%
  \BibitemOpen
  \bibfield  {author} {\bibinfo {author} {\bibfnamefont {G.}~\bibnamefont
  {Carleo}}\ and\ \bibinfo {author} {\bibfnamefont {M.}~\bibnamefont
  {Troyer}},\ }\href@noop {} {\bibfield  {journal} {\bibinfo  {journal}
  {Science}\ }\textbf {\bibinfo {volume} {355}},\ \bibinfo {pages} {602}
  (\bibinfo {year} {2017})},\ \bibinfo {note} {arXiv:1606.02318}\BibitemShut
  {NoStop}%
\bibitem [{\citenamefont {Lopez-Bezanilla}\ and\ \citenamefont {von
  Lilienfeld}(2014)}]{lopez-ml}%
  \BibitemOpen
  \bibfield  {author} {\bibinfo {author} {\bibfnamefont {A.}~\bibnamefont
  {Lopez-Bezanilla}}\ and\ \bibinfo {author} {\bibfnamefont {O.}~\bibnamefont
  {von Lilienfeld}},\ }\href@noop {} {\bibfield  {journal} {\bibinfo  {journal}
  {Phys. Rev. B}\ }\textbf {\bibinfo {volume} {89}} (\bibinfo {year}
  {2014})}\BibitemShut {NoStop}%
\bibitem [{\citenamefont {Schoenholz}\ \emph {et~al.}(2016)\citenamefont
  {Schoenholz}, \citenamefont {Cubuk}, \citenamefont {Sussman},\ and\
  \citenamefont {et~al.}}]{schoenholz-ml}%
  \BibitemOpen
  \bibfield  {author} {\bibinfo {author} {\bibfnamefont {S.}~\bibnamefont
  {Schoenholz}}, \bibinfo {author} {\bibfnamefont {E.}~\bibnamefont {Cubuk}},
  \bibinfo {author} {\bibfnamefont {D.}~\bibnamefont {Sussman}}, \ and\
  \bibinfo {author} {\bibnamefont {et~al.}},\ }\href@noop {} {\bibfield
  {journal} {\bibinfo  {journal} {Nature Physics}\ }\textbf {\bibinfo {volume}
  {12}} (\bibinfo {year} {2016})}\BibitemShut {NoStop}%
\bibitem [{\citenamefont {Wang}(2016)}]{wang-ml}%
  \BibitemOpen
  \bibfield  {author} {\bibinfo {author} {\bibfnamefont {L.}~\bibnamefont
  {Wang}},\ }\href@noop {} {\enquote {\bibinfo {title} {Discovering phase
  transitions with unsupervised learning},}\ } (\bibinfo {year} {2016}),\
  \bibinfo {note} {arXiv: 1606.00318}\BibitemShut {NoStop}%
\bibitem [{\citenamefont {Carrasquilla}\ and\ \citenamefont
  {Melko}(2016)}]{melko-ml}%
  \BibitemOpen
  \bibfield  {author} {\bibinfo {author} {\bibfnamefont {J.}~\bibnamefont
  {Carrasquilla}}\ and\ \bibinfo {author} {\bibfnamefont {R.}~\bibnamefont
  {Melko}},\ }\href@noop {} {\enquote {\bibinfo {title} {Machine learning
  phases of matter},}\ } (\bibinfo {year} {2016}),\ \bibinfo {note} {arXiv:
  1605.01735}\BibitemShut {NoStop}%
\bibitem [{\citenamefont {Mehta}\ and\ \citenamefont
  {Schwab}(2014)}]{ML-prec4}%
  \BibitemOpen
  \bibfield  {author} {\bibinfo {author} {\bibfnamefont {P.}~\bibnamefont
  {Mehta}}\ and\ \bibinfo {author} {\bibfnamefont {D.}~\bibnamefont {Schwab}},\
  }\href@noop {} {\enquote {\bibinfo {title} {An exact mapping between the
  variational renormalization group and deep learning},}\ } (\bibinfo {year}
  {2014}),\ \bibinfo {note} {arXiv: 1410.3831}\BibitemShut {NoStop}%
\bibitem [{\citenamefont {Beny}(2013)}]{ML-prec5}%
  \BibitemOpen
  \bibfield  {author} {\bibinfo {author} {\bibfnamefont {C.}~\bibnamefont
  {Beny}},\ }\href@noop {} {\enquote {\bibinfo {title} {Deep learning and the
  renormalization group},}\ } (\bibinfo {year} {2013}),\ \bibinfo {note}
  {arXiv: 1301.3124}\BibitemShut {NoStop}%
\bibitem [{\citenamefont {Kusne}\ \emph {et~al.}(2014)\citenamefont {Kusne},
  \citenamefont {Gao}, \citenamefont {Mehta},\ and\ \citenamefont
  {et~al.}}]{kusne-ml}%
  \BibitemOpen
  \bibfield  {author} {\bibinfo {author} {\bibfnamefont {A.}~\bibnamefont
  {Kusne}}, \bibinfo {author} {\bibfnamefont {T.}~\bibnamefont {Gao}}, \bibinfo
  {author} {\bibfnamefont {A.}~\bibnamefont {Mehta}}, \ and\ \bibinfo {author}
  {\bibnamefont {et~al.}},\ }\href@noop {} {\bibfield  {journal} {\bibinfo
  {journal} {Scientific Reports}\ }\textbf {\bibinfo {volume} {4}} (\bibinfo
  {year} {2014})}\BibitemShut {NoStop}%
\bibitem [{\citenamefont {Ghiringhelli}\ \emph {et~al.}(2015)\citenamefont
  {Ghiringhelli}, \citenamefont {Vybiral}, \citenamefont {Levchenko},
  \citenamefont {Draxl},\ and\ \citenamefont {Scheffler}}]{ghiringhelli-ml}%
  \BibitemOpen
  \bibfield  {author} {\bibinfo {author} {\bibfnamefont {L.~M.}\ \bibnamefont
  {Ghiringhelli}}, \bibinfo {author} {\bibfnamefont {J.}~\bibnamefont
  {Vybiral}}, \bibinfo {author} {\bibfnamefont {S.~V.}\ \bibnamefont
  {Levchenko}}, \bibinfo {author} {\bibfnamefont {C.}~\bibnamefont {Draxl}}, \
  and\ \bibinfo {author} {\bibfnamefont {M.}~\bibnamefont {Scheffler}},\ }\href
  {\doibase 10.1103/PhysRevLett.114.105503} {\bibfield  {journal} {\bibinfo
  {journal} {Phys. Rev. Lett.}\ }\textbf {\bibinfo {volume} {114}},\ \bibinfo
  {pages} {105503} (\bibinfo {year} {2015})}\BibitemShut {NoStop}%
\bibitem [{\citenamefont {Kalinin}\ \emph {et~al.}(2015)\citenamefont
  {Kalinin}, \citenamefont {Sumpter},\ and\ \citenamefont
  {Archibald}}]{kalinin-ml}%
  \BibitemOpen
  \bibfield  {author} {\bibinfo {author} {\bibfnamefont {S.}~\bibnamefont
  {Kalinin}}, \bibinfo {author} {\bibfnamefont {B.}~\bibnamefont {Sumpter}}, \
  and\ \bibinfo {author} {\bibfnamefont {R.}~\bibnamefont {Archibald}},\
  }\href@noop {} {\bibfield  {journal} {\bibinfo  {journal} {Nature Materials}\
  }\textbf {\bibinfo {volume} {14}} (\bibinfo {year} {2015})}\BibitemShut
  {NoStop}%
\bibitem [{\citenamefont {Arsenault}\ \emph {et~al.}(2014)\citenamefont
  {Arsenault}, \citenamefont {Lopez-Bezanilla}, \citenamefont {von
  Lilienfeld},\ and\ \citenamefont {Millis}}]{millis-prb}%
  \BibitemOpen
  \bibfield  {author} {\bibinfo {author} {\bibfnamefont {L.-F.}\ \bibnamefont
  {Arsenault}}, \bibinfo {author} {\bibfnamefont {A.}~\bibnamefont
  {Lopez-Bezanilla}}, \bibinfo {author} {\bibfnamefont {O.~A.}\ \bibnamefont
  {von Lilienfeld}}, \ and\ \bibinfo {author} {\bibfnamefont {A.~J.}\
  \bibnamefont {Millis}},\ }\href {\doibase 10.1103/PhysRevB.90.155136}
  {\bibfield  {journal} {\bibinfo  {journal} {Phys. Rev. B}\ }\textbf {\bibinfo
  {volume} {90}},\ \bibinfo {pages} {155136} (\bibinfo {year}
  {2014})}\BibitemShut {NoStop}%
\bibitem [{\citenamefont {Arsenault}\ \emph {et~al.}(2015)\citenamefont
  {Arsenault}, \citenamefont {{von~Lilienfeld}},\ and\ \citenamefont
  {Millis}}]{millis-arxiv}%
  \BibitemOpen
  \bibfield  {author} {\bibinfo {author} {\bibfnamefont {L.-F.}\ \bibnamefont
  {Arsenault}}, \bibinfo {author} {\bibfnamefont {O.~A.}\ \bibnamefont
  {{von~Lilienfeld}}}, \ and\ \bibinfo {author} {\bibfnamefont {A.~J.}\
  \bibnamefont {Millis}},\ }\href@noop {} {\enquote {\bibinfo {title} {Machine
  learning for many-body physics: efficient solution of dynamical mean-field
  theory},}\ } (\bibinfo {year} {2015}),\ \bibinfo {note}
  {arXiv:1506.08858}\BibitemShut {NoStop}%
\bibitem [{\citenamefont {Fisher}(1974)}]{fisher-rmp}%
  \BibitemOpen
  \bibfield  {author} {\bibinfo {author} {\bibfnamefont {M.~E.}\ \bibnamefont
  {Fisher}},\ }\href {\doibase 10.1103/RevModPhys.46.597} {\bibfield  {journal}
  {\bibinfo  {journal} {Rev. Mod. Phys.}\ }\textbf {\bibinfo {volume} {46}},\
  \bibinfo {pages} {597} (\bibinfo {year} {1974})}\BibitemShut {NoStop}%
\bibitem [{\citenamefont {Stanley}(1971)}]{stanley-book}%
  \BibitemOpen
  \bibfield  {author} {\bibinfo {author} {\bibfnamefont {H.~E.}\ \bibnamefont
  {Stanley}},\ }\href@noop {} {\emph {\bibinfo {title} {Introduction to Phase
  Transitions and Critical Phenomena}}}\ (\bibinfo  {publisher} {Oxford
  University Press},\ \bibinfo {year} {1971})\BibitemShut {NoStop}%
\bibitem [{\citenamefont {Moller}(1993)}]{moller-scg}%
  \BibitemOpen
  \bibfield  {author} {\bibinfo {author} {\bibfnamefont {M.~F.}\ \bibnamefont
  {Moller}},\ }\href@noop {} {\bibfield  {journal} {\bibinfo  {journal} {Neural
  Networks}\ }\textbf {\bibinfo {volume} {6}},\ \bibinfo {pages} {525}
  (\bibinfo {year} {1993})}\BibitemShut {NoStop}%
\bibitem [{\citenamefont {Towns}\ \emph {et~al.}(2014)\citenamefont {Towns},
  \citenamefont {Cockerill}, \citenamefont {Dahan}, \citenamefont {Foster},
  \citenamefont {Gaither}, \citenamefont {Grimshaw}, \citenamefont {Hazlewood},
  \citenamefont {Lathrop}, \citenamefont {Lifka}, \citenamefont {Peterson},
  \citenamefont {Roskies}, \citenamefont {Scott},\ and\ \citenamefont
  {Wilkins-Diehr}}]{XSEDE}%
  \BibitemOpen
  \bibfield  {author} {\bibinfo {author} {\bibfnamefont {J.}~\bibnamefont
  {Towns}}, \bibinfo {author} {\bibfnamefont {T.}~\bibnamefont {Cockerill}},
  \bibinfo {author} {\bibfnamefont {M.}~\bibnamefont {Dahan}}, \bibinfo
  {author} {\bibfnamefont {I.}~\bibnamefont {Foster}}, \bibinfo {author}
  {\bibfnamefont {K.}~\bibnamefont {Gaither}}, \bibinfo {author} {\bibfnamefont
  {A.}~\bibnamefont {Grimshaw}}, \bibinfo {author} {\bibfnamefont
  {V.}~\bibnamefont {Hazlewood}}, \bibinfo {author} {\bibfnamefont
  {S.}~\bibnamefont {Lathrop}}, \bibinfo {author} {\bibfnamefont
  {D.}~\bibnamefont {Lifka}}, \bibinfo {author} {\bibfnamefont {G.~D.}\
  \bibnamefont {Peterson}}, \bibinfo {author} {\bibfnamefont {R.}~\bibnamefont
  {Roskies}}, \bibinfo {author} {\bibfnamefont {J.~R.}\ \bibnamefont {Scott}},
  \ and\ \bibinfo {author} {\bibfnamefont {N.}~\bibnamefont {Wilkins-Diehr}},\
  }\href {\doibase doi.ieeecomputersociety.org/10.1109/MCSE.2014.80} {\bibfield
   {journal} {\bibinfo  {journal} {Computing in Science and Engineering}\
  }\textbf {\bibinfo {volume} {16}},\ \bibinfo {pages} {62} (\bibinfo {year}
  {2014})}\BibitemShut {NoStop}%
\bibitem [{\citenamefont {Lupi}\ \emph {et~al.}(2010)\citenamefont {Lupi},
  \citenamefont {Baldassarre}, \citenamefont {Mansart}, \citenamefont
  {Perucchi}, \citenamefont {Barinov}, \citenamefont {Dudin}, \citenamefont
  {Papalazarou}, \citenamefont {Rodolakis}, \citenamefont {Rueff},
  \citenamefont {Iti{\'e}}, \citenamefont {Ravy}, \citenamefont {Nicoletti},
  \citenamefont {Postorino}, \citenamefont {Hansmann}, \citenamefont {Parragh},
  \citenamefont {Toschi}, \citenamefont {Saha-Dasgupta}, \citenamefont
  {Andersen}, \citenamefont {Sangiovanni}, \citenamefont {Held},\ and\
  \citenamefont {Marsi}}]{V2O3}%
  \BibitemOpen
  \bibfield  {author} {\bibinfo {author} {\bibfnamefont {S.}~\bibnamefont
  {Lupi}}, \bibinfo {author} {\bibfnamefont {L.}~\bibnamefont {Baldassarre}},
  \bibinfo {author} {\bibfnamefont {B.}~\bibnamefont {Mansart}}, \bibinfo
  {author} {\bibfnamefont {A.}~\bibnamefont {Perucchi}}, \bibinfo {author}
  {\bibfnamefont {A.}~\bibnamefont {Barinov}}, \bibinfo {author} {\bibfnamefont
  {P.}~\bibnamefont {Dudin}}, \bibinfo {author} {\bibfnamefont
  {E.}~\bibnamefont {Papalazarou}}, \bibinfo {author} {\bibfnamefont
  {F.}~\bibnamefont {Rodolakis}}, \bibinfo {author} {\bibfnamefont {J.~P.}\
  \bibnamefont {Rueff}}, \bibinfo {author} {\bibfnamefont {J.~P.}\ \bibnamefont
  {Iti{\'e}}}, \bibinfo {author} {\bibfnamefont {S.}~\bibnamefont {Ravy}},
  \bibinfo {author} {\bibfnamefont {D.}~\bibnamefont {Nicoletti}}, \bibinfo
  {author} {\bibfnamefont {P.}~\bibnamefont {Postorino}}, \bibinfo {author}
  {\bibfnamefont {P.}~\bibnamefont {Hansmann}}, \bibinfo {author}
  {\bibfnamefont {N.}~\bibnamefont {Parragh}}, \bibinfo {author} {\bibfnamefont
  {A.}~\bibnamefont {Toschi}}, \bibinfo {author} {\bibfnamefont
  {T.}~\bibnamefont {Saha-Dasgupta}}, \bibinfo {author} {\bibfnamefont {O.~K.}\
  \bibnamefont {Andersen}}, \bibinfo {author} {\bibfnamefont {G.}~\bibnamefont
  {Sangiovanni}}, \bibinfo {author} {\bibfnamefont {K.}~\bibnamefont {Held}}, \
  and\ \bibinfo {author} {\bibfnamefont {M.}~\bibnamefont {Marsi}},\
  }\href@noop {} {\bibfield  {journal} {\bibinfo  {journal} {Nature
  Communications}\ }\textbf {\bibinfo {volume} {1}},\ \bibinfo {pages} {105}
  (\bibinfo {year} {2010})}\BibitemShut {NoStop}%
\bibitem [{\citenamefont {Battisti}\ \emph {et~al.}(2016)\citenamefont
  {Battisti}, \citenamefont {Bastiaans}, \citenamefont {Fedoseev},
  \citenamefont {de~la Torre}, \citenamefont {Iliopoulos}, \citenamefont
  {Tamai}, \citenamefont {Hunter}, \citenamefont {Perry}, \citenamefont
  {Zaanen}, \citenamefont {Baumberger},\ and\ \citenamefont
  {Allan}}]{milan-iridates}%
  \BibitemOpen
  \bibfield  {author} {\bibinfo {author} {\bibfnamefont {I.}~\bibnamefont
  {Battisti}}, \bibinfo {author} {\bibfnamefont {K.~M.}\ \bibnamefont
  {Bastiaans}}, \bibinfo {author} {\bibfnamefont {V.}~\bibnamefont {Fedoseev}},
  \bibinfo {author} {\bibfnamefont {A.}~\bibnamefont {de~la Torre}}, \bibinfo
  {author} {\bibfnamefont {N.}~\bibnamefont {Iliopoulos}}, \bibinfo {author}
  {\bibfnamefont {A.}~\bibnamefont {Tamai}}, \bibinfo {author} {\bibfnamefont
  {E.~C.}\ \bibnamefont {Hunter}}, \bibinfo {author} {\bibfnamefont {R.~S.}\
  \bibnamefont {Perry}}, \bibinfo {author} {\bibfnamefont {J.}~\bibnamefont
  {Zaanen}}, \bibinfo {author} {\bibfnamefont {F.}~\bibnamefont {Baumberger}},
  \ and\ \bibinfo {author} {\bibfnamefont {M.~P.}\ \bibnamefont {Allan}},\
  }\href@noop {} {\bibfield  {journal} {\bibinfo  {journal} {Nature Physics}\ }
  (\bibinfo {year} {2016})}\BibitemShut {NoStop}%
\end{thebibliography}

%

\end{document}